\def\year{2020}\relax
\documentclass[letterpaper]{article} 
\usepackage{aaai20arxiv}  
\usepackage{times}  
\usepackage{helvet} 
\usepackage{courier}  
\usepackage[hyphens]{url}  
\usepackage{graphicx} 
\usepackage{graphicx}  
\newtheorem{definition}{Definition}
\newtheorem{example}{Example}
\usepackage{braket}
\usepackage{tikz}
\usetikzlibrary{decorations.pathreplacing,arrows,quotes,angles}
\newcommand\C{{\mathbf C}}
\newcommand\psiangle{18}
\title{Quantum Weighted Model Counting}
\author{Fabrizio Riguzzi\\ 
Department of Mathematics and Computer Science, University of Ferrara\\
Via Saragat 1, 44122, Ferrara Italy\\
fabrizio.riguzzi@unife.it}

\begin{document}

\maketitle

\begin{abstract}
In Weighted Model Counting (WMC) we assign weights to Boolean literals and we want to compute the sum of the weights of the models of a Boolean function where the weight of a model is the product of the weights of its literals.

WMC was shown to be particularly effective for performing inference in graphical models, with a complexity of $O(n2^w)$ where $n$ is the number of variables and $w$ is the treewidth.

In this paper, we propose a quantum algorithm for performing WMC, Quantum WMC (QWMC), that 
modifies the quantum model counting algorithm to take into account the weights. In turn, the model counting algorithm uses the algorithms of quantum search,  phase estimation and Fourier transform.

In the black box model of computation, where we can only query an oracle for evaluating the Boolean function given an assignment,
QWMC solves the problem approximately with
a complexity of $\Theta(2^{\frac{n}{2}})$ oracle calls while classically the best complexity is  $\Theta(2^n)$, thus achieving a quadratic speedup. 

\end{abstract}

\section{Introduction}
Weighted Model Counting (WMC) is the problem of computing the sum of the weights of the models of 
a propositional formula, where the weight of a model is given by multiplying the weights of the literals in 
it. WMC proved effective for performing inference in graphical models \cite{DBLP:journals/ai/ChaviraD08,DBLP:conf/aaai/SangBK05}. While other graphical model inference algorithms \cite{lauritzen1988local,DBLP:journals/jair/ZhangP96,dechter1999bucket,darwiche2001recursive}
take time $\Theta(n2^w)$  where $n$ is the number of variables
and $w$ is the treewidth of the network, WMC takes time $O(n2^w)$, i.e., exponential in the treewidth
in the worst case \cite{DBLP:journals/ai/ChaviraD08}. WMC does so by exploiting structure in the graphical model in the form of context-specific independence and determinism.

In this paper we propose to perform WMC  using a quantum computer, i.e.,  Quantum WMC (QWMC). Quantum computing  \cite{nielsen2010quantum} is the use of quantum mechanics to perform 
computation. Various algorithms have been proposed for quantum computers that improve over
their classical counterpart, the most prominent are: Shor's algorithm \cite{shor}, that factorizes integers in polynomial time while no classical polynomial algorithm is known, and quantum search, that has a quadratic speedup over classical search \cite{Grover:1996:FQM:237814.237866,grover1996fast,grover1997quantum}.

To perform QWMC, we use various quantum  algorithms. In particular, we adapt the method 
of quantum model counting \cite{boyer1998tight,DBLP:conf/icalp/BrassardHT98}. to take into account weights. Quantum model counting in turn is based on
quantum search using Grover's algorithm \cite{Grover:1996:FQM:237814.237866,grover1996fast,grover1997quantum} and on quantum phase estimation  \cite{cleve1998quantum}, the latter using  quantum Fourier transform \cite{coppersmith2002approximate}.

Here we consider the problem of WMC under a black box computation model where we don't know 
anything about the propositional formula, we only have the possibility of querying an oracle giving the
 value of the formula for an assignment
of the propositional variables, and we consider the complexity in terms of oracle calls. In this computation model, QWMC solves the problem approximately with
a complexity of $\Theta(2^{\frac{n}{2}})$ while classically the best complexity is  $\Theta(2^n)$, thus achieving a quadratic speedup. 

QWMC may be useful for models with high treewidth: if the treewidth is larger than half  
the number of variables, then QWMC performs better than other inference algorithms.

\section{Weighted Model Counting}
Propositional satisfiability (SAT) is the problem of deciding
whether a logical formula over Boolean variables evaluates to true for some truth value assignment
of the Boolean variables. If an assignment $M$ makes formula $\phi$ true we write $M\models \phi$.
Model counting or \#SAT \cite{DBLP:series/faia/GomesSS09} aims at computing
the number of satisfying assignments of a propositional sentence.

Weighted model counting (WMC) \cite{DBLP:journals/ai/ChaviraD08} generalizes model counting  by giving each assignment a weight
and aiming at computing the sum of the weights of all satisfying
assignments. 

\begin{definition}
Given a formula $\phi$ in propositional logic over
literals $L$ (Boolean variables or their negation), and a weight function $w : L \to R^{\geq 0 }$, the weighted
model count (WMC) is defined as:
$$WMC(\phi,w) =\sum_{M\models \phi} weight(M,w)$$
where
$weight(M,w) =\prod_{l\in M}w(l)$
\end{definition}

\begin{example}\label{sprinklerb}
Let us consider an example inspired by the sprinkler problem of \cite{pearl88}: we have three Boolean variable, $s$, $r$, $w$ representing respectively propositions ``the sprinkler was on'', `ìt rained last night'' and ``the grass is wet''. We know that if the sprinkler was on the grass is wet ($s\rightarrow w$), if it rained last night the grass is wet ($r\rightarrow w$) and that the the sprinkler being on and rain last night cannot be true at the same time ($s,r\rightarrow$). Transforming the formula into conjunctive normal formal we obtain the formula
$$\phi=(\neg s \vee w)\wedge(\neg r \vee w)\wedge(\neg s\vee \neg r)$$
Suppose the weights of literals are
$w(s)=0.3$, $w(\neg s)=0.7$, $w(r)=0.2$, $w(\neg r)=0.8$, $w(w)=0.5$ and $w(\neg w)=0.5$, Table \ref{sprinkler_table} shows the worlds of $\phi$ together with the weight of each world.
The WMC of $\phi$ is thus
$WMC(\phi,w)=0.28+0.28+0.07+0.12=0.75$
\end{example}
\begin{table}
\centering
\begin{tabular}{|ccc|c|c|}
\hline
s&r&w&$\phi$&W\\
\hline
0&0&0&1&$0.7\cdot 0.8\cdot 0.5=0.28$\\
0&0&1&1&$0.7\cdot 0.8\cdot 0.5=0.28$\\
0&1&0&0&$0.7\cdot 0.2\cdot 0.5=0.07$\\
0&1&1&1&$0.7\cdot 0.2\cdot 0.5=0.07$\\
1&0&0&0&$0.3\cdot 0.8\cdot 0.5=0.12$\\
1&0&1&1&$0.3\cdot 0.8\cdot 0.5=0.12$\\
1&1&0&0&$0.3\cdot 0.2\cdot 0.5=0.03$\\
1&1&1&0&$0.3\cdot 0.2\cdot 0.5=0.03$\\
\hline
\end{tabular}

\caption{Worlds for formula $\phi$ of Example \ref{sprinklerb}.\label{sprinkler_table}}
\end{table}

%

\section{Quantum Computing}

Here we provide a brief introduction to quantum computing following \cite{nielsen2010quantum}.
As the bit is at the basis of classical computing, the \emph{quantum bit} or \emph{qubit} is at the basis of quantum computing. A qubit is a mathematical object that can have various physical implementations.
Mathematically it is a unit vector in the $\C^2$ space  where $\C$ is the set of complex numbers.
A bit can be in one of two states, similarly a qubit has a state which is its vector in  $\C^2$.
Usually qubit are represented using the Dirac notation where $\ket{\psi}$ is a two dimensional column vector representing the state of a qubit while $\bra{\psi}$ is a two dimensional row vector. Usually, the special states $\ket{0}$ and $\ket{1}$ are identified:
they are called computational basis states and form an orthonormal basis for $\C^2$.
Any qubit state $\ket{\psi}$ can be expressed as a linear combination of the computational basis states:
$$\ket{\psi}=\alpha\ket{0}+\beta\ket{1}=\left[\begin{array}{c}\alpha\\\beta\end{array}\right]$$
where $\alpha$ and $\beta$ are complex number such that $|\alpha|^2+|\beta|^2=1$.
In this case we say that $\ket{\psi}$ is in a superposition of states $\ket{0}$ and $\ket{1}$.

In this paper we follow the quantum circuit model of computation where each qubit corresponds to a wire and quantum gates are applied to sets of wires.

\begin{figure}
\centering
\begin{tikzpicture}[scale=1.000000,x=1pt,y=1pt]
\filldraw[color=white] (0.000000, -7.500000) rectangle (24.000000, 7.500000);
\draw[color=black] (0.000000,0.000000) -- (24.000000,0.000000);
\begin{scope}
\draw[fill=white] (12.000000, -0.000000) +(-45.000000:8.485281pt and 8.485281pt) -- +(45.000000:8.485281pt and 8.485281pt) -- +(135.000000:8.485281pt and 8.485281pt) -- +(225.000000:8.485281pt and 8.485281pt) -- cycle;
\clip (12.000000, -0.000000) +(-45.000000:8.485281pt and 8.485281pt) -- +(45.000000:8.485281pt and 8.485281pt) -- +(135.000000:8.485281pt and 8.485281pt) -- +(225.000000:8.485281pt and 8.485281pt) -- cycle;
\draw (12.000000, -0.000000) node {$X$};
\end{scope}
\end{tikzpicture}\hspace{0.5cm}\begin{tikzpicture}[scale=1.000000,x=1pt,y=1pt]
\filldraw[color=white] (0.000000, -7.500000) rectangle (24.000000, 7.500000);
\draw[color=black] (0.000000,0.000000) -- (24.000000,0.000000);
\begin{scope}
\draw[fill=white] (12.000000, -0.000000) +(-45.000000:8.485281pt and 8.485281pt) -- +(45.000000:8.485281pt and 8.485281pt) -- +(135.000000:8.485281pt and 8.485281pt) -- +(225.000000:8.485281pt and 8.485281pt) -- cycle;
\clip (12.000000, -0.000000) +(-45.000000:8.485281pt and 8.485281pt) -- +(45.000000:8.485281pt and 8.485281pt) -- +(135.000000:8.485281pt and 8.485281pt) -- +(225.000000:8.485281pt and 8.485281pt) -- cycle;
\draw (12.000000, -0.000000) node {$H$};
\end{scope}
\end{tikzpicture}\hspace{0.5cm}\begin{tikzpicture}[scale=1.000000,x=1pt,y=1pt]
\filldraw[color=white] (0.000000, -7.500000) rectangle (42.000000, 7.500000);
\draw[color=black] (0.000000,0.000000) -- (42.000000,0.000000);
\begin{scope}
\draw[fill=white] (21.000000, -0.000000) +(-45.000000:21.213203pt and 8.485281pt) -- +(45.000000:21.213203pt and 8.485281pt) -- +(135.000000:21.213203pt and 8.485281pt) -- +(225.000000:21.213203pt and 8.485281pt) -- cycle;
\clip (21.000000, -0.000000) +(-45.000000:21.213203pt and 8.485281pt) -- +(45.000000:21.213203pt and 8.485281pt) -- +(135.000000:21.213203pt and 8.485281pt) -- +(225.000000:21.213203pt and 8.485281pt) -- cycle;
\draw (21.000000, -0.000000) node {$R_y(\theta)$};
\end{scope}
\end{tikzpicture}

\begin{tikzpicture}[scale=1.000000,x=1pt,y=1pt]
\filldraw[color=white] (0.000000, -7.500000) rectangle (24.000000, 7.500000);
\draw[color=black] (0.000000,0.000000) -- (12.000000,0.000000);
\draw[color=black] (12.000000,-0.500000) -- (24.000000,-0.500000);
\draw[color=black] (12.000000,0.500000) -- (24.000000,0.500000);
\draw[fill=white] (6.000000, -6.000000) rectangle (18.000000, 6.000000);
\draw[very thin] (12.000000, 0.600000) arc (90:150:6.000000pt);
\draw[very thin] (12.000000, 0.600000) arc (90:30:6.000000pt);
\draw[->,>=stealth] (12.000000, -5.400000) -- +(80:10.392305pt);
\end{tikzpicture}\hspace{0.5cm}\begin{tikzpicture}[scale=1.000000,x=1pt,y=1pt]
\filldraw[color=white] (0.000000, -7.500000) rectangle (18.000000, 22.500000);
\draw[color=black] (0.000000,15.000000) -- (18.000000,15.000000);
\draw[color=black] (0.000000,15.000000) node[left] {$\ket{a}$};
\draw[color=black] (0.000000,0.000000) -- (18.000000,0.000000);
\draw[color=black] (0.000000,0.000000) node[left] {$\ket{b}$};
\draw (9.000000,15.000000) -- (9.000000,0.000000);
\begin{scope}
\draw[fill=white] (9.000000, 0.000000) circle(3.000000pt);
\clip (9.000000, 0.000000) circle(3.000000pt);
\draw (6.000000, 0.000000) -- (12.000000, 0.000000);
\draw (9.000000, -3.000000) -- (9.000000, 3.000000);
\end{scope}
\filldraw (9.000000, 15.000000) circle(1.500000pt);
\draw[color=black] (18.000000,15.000000) node[right] {$\ket{a}$};
\draw[color=black] (18.000000,0.000000) node[right] {$\ket{b\oplus a}$};
\end{tikzpicture}
\caption{Examples of quantum gates.}
\label{xgate}
\end{figure}
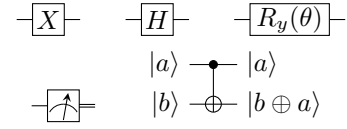

Quantum gates are represented by matrices with complex elements.
The \emph{adjoint} or \emph{Hermitian conjugate} of a matrix $M$, denoted by $M^\dagger$, is the conjugate and transpose matrix $M^\dagger=(M^*)^T$. A matrix is \emph{unitary} if $M^\dagger M=I$.
Quantum gates are represented by unitary matrices. 
The simplest gates are those operating on a single qubit and belong to $\C^{2\times 2}$.
For  example, the counterpart of the NOT Boolean gate for classical bits is $X$ defined as
$$X=\left[\begin{array}{cc}0&1\\
1&0\end{array}\right]$$
and represented as in Figure \ref{xgate} top left.
Another important gate is the \emph{Hadamard gate} (see Figure \ref{xgate} top center)
$$H=\frac{1}{\sqrt{2}}\left[\begin{array}{cc}1&1\\
1&-1\end{array}\right].$$
A gate that we will use in the following is:
$$R_y(\theta)=\left[\begin{array}{cc}
\cos\frac{\theta}{2}&-\sin\frac{\theta}{2}\\
\sin\frac{\theta}{2}&\cos\frac{\theta}{2}
\end{array}\right]
$$
that applies a rotation of $\theta/2$ radians, with $\theta$ user defined, see Figure \ref{xgate} top right.

Another operation we can apply to a qubit is \emph{measurement}. There are various types of measurements, here we consider only the one with respect to the computational basis that, given a qubit $\alpha\ket{0}+\beta\ket{1}$, returns a classical bit, namely 0 with probability $|\alpha|^2$ and 1 with probability $|\beta|^2$. Since qubits are unit vectors, this operation is well-defined. Measurement is represented as in Figure \ref{xgate} bottom left.

When we have more than one bit, we have a composite physical system and the state space expands accordingly: for $n$ qubits, there are $2^n$ computational basis states, e.g., if $n=2$ the basis states are $\ket{00}$, $\ket{01}$, $\ket{10}$ and
$\ket{11}$ and the state of the qubits can be written as
$$\ket{\psi}=\alpha_{00}\ket{00}+\alpha_{01}\ket{01}+\alpha_{10}\ket{10}+\alpha_{11}\ket{11}$$
Moreover, the state space of a composite physical system is the tensor product
of the state spaces of the component physical systems.

The tensor product of two column vectors $a$ and $b$ is $ab^T$.
So the tensor product of two qubits 

$$\ket{a}=a_0\ket{0}+a_1\ket{1}=\left[\begin{array}{c}
a_0\\
a_1
\end{array}\right]$$
$$
\ket{b}=b_0\ket{0}+b_1\ket{1}=\left[\begin{array}{c}
b_0\\
b_1
\end{array}\right]
$$
is
$$\ket{a}\otimes \ket{b}=\left[\begin{array}{c}
a_0b_0\\
a_0b_1\\
a_1b_0\\
a_1b_1\\
\end{array}\right]=$$
$$a_0b_0\ket{00}+
a_0b_1\ket{01}+
a_1b_0\ket{10}+
a_1b_1\ket{11}
$$
For two qubits, the most important gate is the \emph{controlled-NOT} or \emph{CNOT} gate that has two inputs, the 
control and the target qubits, and acts by flipping the target qubit if the control bit is set to 1 and does nothing if the control bit is set to 0. It can also be defined as a gate that  operates as $\ket{ab}\to\ket{a,b\oplus a}$ where $\oplus$ is the XOR operation, see Figure \ref{xgate} bottom right.

Any multiple qubit
logic gate may be composed from CNOT and single qubit gates.

 CNOT may be generalized to the case of more than two bits: in this case, the extra qubits act as controls and the target is flipped if all controls are 1. Moreover, given an operator $U$, it is possible to define a control-$U$ operator defined as $\ket{ab}\to\ket{a,U^ab}$: if $a=0$ it does nothing, otherwise it applies operator $U$ to $b$.

\begin{figure}[t]
\centering
\begin{footnotesize}
\begin{tikzpicture}[scale=0.800000,x=1pt,y=1pt]
\filldraw[color=white] (0.000000, -7.500000) rectangle (96.000000, 97.500000);
\draw[color=black] (0.000000,90.000000) -- (96.000000,90.000000);
\draw[color=black] (0.000000,90.000000) node[left] {$\ket{s}$};
\draw[color=black] (0.000000,75.000000) -- (96.000000,75.000000);
\draw[color=black] (0.000000,75.000000) node[left] {$\ket{r}$};
\draw[color=black] (0.000000,60.000000) -- (96.000000,60.000000);
\draw[color=black] (0.000000,60.000000) node[left] {$\ket{w}$};
\draw[color=black] (0.000000,45.000000) -- (96.000000,45.000000);
\draw[color=black] (0.000000,45.000000) node[left] {$\ket{1}$};
\draw[color=black] (0.000000,30.000000) -- (96.000000,30.000000);
\draw[color=black] (0.000000,30.000000) node[left] {$\ket{1}$};
\draw[color=black] (0.000000,15.000000) -- (96.000000,15.000000);
\draw[color=black] (0.000000,15.000000) node[left] {$\ket{1}$};
\draw[color=black] (0.000000,0.000000) -- (96.000000,0.000000);
\draw[color=black] (0.000000,0.000000) node[left] {$\ket{0}$};
\begin{scope}
\draw[fill=white] (12.000000, 60.000000) +(-45.000000:8.485281pt and 8.485281pt) -- +(45.000000:8.485281pt and 8.485281pt) -- +(135.000000:8.485281pt and 8.485281pt) -- +(225.000000:8.485281pt and 8.485281pt) -- cycle;
\clip (12.000000, 60.000000) +(-45.000000:8.485281pt and 8.485281pt) -- +(45.000000:8.485281pt and 8.485281pt) -- +(135.000000:8.485281pt and 8.485281pt) -- +(225.000000:8.485281pt and 8.485281pt) -- cycle;
\draw (12.000000, 60.000000) node {$X$};
\end{scope}
\draw (33.000000,90.000000) -- (33.000000,45.000000);
\begin{scope}
\draw[fill=white] (33.000000, 45.000000) circle(3.000000pt);
\clip (33.000000, 45.000000) circle(3.000000pt);
\draw (30.000000, 45.000000) -- (36.000000, 45.000000);
\draw (33.000000, 42.000000) -- (33.000000, 48.000000);
\end{scope}
\filldraw (33.000000, 90.000000) circle(1.500000pt);
\filldraw (33.000000, 75.000000) circle(1.500000pt);
\draw (51.000000,75.000000) -- (51.000000,30.000000);
\begin{scope}
\draw[fill=white] (51.000000, 30.000000) circle(3.000000pt);
\clip (51.000000, 30.000000) circle(3.000000pt);
\draw (48.000000, 30.000000) -- (54.000000, 30.000000);
\draw (51.000000, 27.000000) -- (51.000000, 33.000000);
\end{scope}
\filldraw (51.000000, 75.000000) circle(1.500000pt);
\filldraw (51.000000, 60.000000) circle(1.500000pt);
\draw (69.000000,90.000000) -- (69.000000,15.000000);
\begin{scope}
\draw[fill=white] (69.000000, 15.000000) circle(3.000000pt);
\clip (69.000000, 15.000000) circle(3.000000pt);
\draw (66.000000, 15.000000) -- (72.000000, 15.000000);
\draw (69.000000, 12.000000) -- (69.000000, 18.000000);
\end{scope}
\filldraw (69.000000, 90.000000) circle(1.500000pt);
\filldraw (69.000000, 60.000000) circle(1.500000pt);
\draw (87.000000,45.000000) -- (87.000000,0.000000);
\begin{scope}
\draw[fill=white] (87.000000, 0.000000) circle(3.000000pt);
\clip (87.000000, 0.000000) circle(3.000000pt);
\draw (84.000000, 0.000000) -- (90.000000, 0.000000);
\draw (87.000000, -3.000000) -- (87.000000, 3.000000);
\end{scope}
\filldraw (87.000000, 45.000000) circle(1.500000pt);
\filldraw (87.000000, 30.000000) circle(1.500000pt);
\filldraw (87.000000, 15.000000) circle(1.500000pt);
\draw[color=black] (96.000000,90.000000) node[right] {$\ket{s}$};
\draw[color=black] (96.000000,75.000000) node[right] {$\ket{r}$};
\draw[color=black] (96.000000,60.000000) node[right] {$\ket{\neg w}$};
\draw[color=black] (96.000000,45.000000) node[right] {$\ket{\neg (s \wedge r)}=\ket{\neg s \vee \neg r}$};
\draw[color=black] (96.000000,30.000000) node[right] {$\ket{\neg (s \wedge \neg w)}=\ket{\neg s\vee w}$};
\draw[color=black] (96.000000,15.000000) node[right] {$\ket{\neg (r \wedge \neg w)}=\ket{\neg r\vee w}$};
\draw[color=black] (96.000000,0.000000) node[right] {$\ket{(\neg s \vee \neg r)\wedge (\neg s\vee w)\wedge (\neg r\vee w)}$};
\end{tikzpicture}
\end{footnotesize}

\caption{Quantum circuit for computing $\phi$}
\label{sprinklerq}
\end{figure}
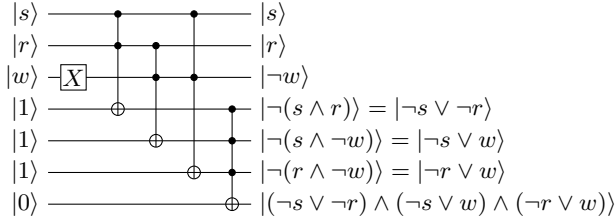

\begin{example}
The quantum circuit for computing the value of formula $\phi$ from Example \ref{sprinklerb}
is shown in Figure \ref{sprinklerq}. 
\end{example}

Quantum circuits should be read from left to right. Each line or wire correspond to a qubit and starts
in a computational basis state, usually $\ket{0}$ unless otherwise indicated.
The circuit in Figure \ref{sprinklerq} contains one wire for each Boolean variable of Example \ref{sprinklerb} plus four other wires that represent the so called \emph{ancilla qubits}. 
Ancilla qubits are used in order to make the circuit reversible. The bottom ancilla qubit contains the truth value of function $\phi$.

%

\section{Quantum Fourier Transform}

The discrete Fourier transform computer a vector of complex numbers $y_0,\ldots y_{N-1}$ given
a vector of complex numbers $x_0,\ldots, x_{N-1}$ as follows
$$y_k=\frac{1}{\sqrt{N}}\sum_{j=0}^{N-1}x_je^{2\pi ijk/N}$$
The \emph{quantum Fourier transform} \cite{coppersmith2002approximate} is similar, it takes an orthonormal basis $\ket{0},\ldots,\ket{N-1}$ 
and transforms it as:
$$\ket{j}\to \frac{1}{\sqrt{N}}\sum_{k=0}^{N-1}e^{2\pi i j k/N}\ket{k}$$
It is a Fourier transform because the action on an arbitrary state is
$$\sum_{j=0}^{N-1}x_j\ket{j}\to\sum_{k=0}^{N-1}y_k\ket{k}$$
with $y_k$ as in the discrete Fourier transform.

The quantum Fourier transform can be given a \emph{product representation} \cite{cleve1998quantum,griffiths1996semiclassical}:
\begin{tiny}
\begin{equation}
\label{product_rep}
\begin{array}{l}\ket{j_1,\ldots,j_n}\to\\
 \frac{\left(\ket{0}+e^{2\pi 0.j_n}\ket{1}\right)\left(\ket{0}+e^{2\pi 0.j_{n-1}j_n}\ket{1}\right)\cdots\left(\ket{0}+e^{2\pi 0.j_1j_2\cdots j_n}\ket{1}\right)}{2^{n/2}}
 \end{array}
 \end{equation}
\end{tiny}
where we assumed that $N=2^n$, the state $\ket{j}$ is written using the binary representation
$j=j_1j_2\ldots j_n$ and $0.j_lj_{l+1}\ldots j_m$ represents the number $j_l/2+j_{l+1}/4+\ldots+j_m/2^{m-l+1}$.
The quantum Fourier transform requires $\Theta(n^2)$ gates.

\section{Quantum Phase Estimation}
In the problem of \emph{quantum phase estimation} \cite{cleve1998quantum}, we are given an operator $U$ and one of its eigenvectors
$\ket{u}$ with eigenvalue $e^{2\pi i\varphi}$ and we want to find the value of $\varphi$.
We assume that that we have black boxes that can prepare the state $\ket{u}$ and perform
controlled-$U^{2^j}$ operations for non negative integers $j$.

Phase estimation uses two registers, one with $t$ qubits initially in state $\ket{0}$ and the other with as many qubits as are necessary to store $\ket{u}$ that is also its initial state.

The first stage of phase estimation is shown in Figure \ref{phase_est}. If the phase can be represented with exactly $t$ bits as $\varphi=0.\varphi_1\ldots\varphi_t$, the first stage brings the first register to state
\begin{tiny}
$$
 \frac{\left(\ket{0}+e^{2\pi 0.\varphi_t}\ket{1}\right)\left(\ket{0}+e^{2\pi 0.\varphi_{t-1}\varphi_t}\ket{1}\right)\cdots\left(\ket{0}+e^{2\pi 0.\varphi_1\cdots \varphi_t}\ket{1}\right)}{2^{n/2}}
$$
\end{tiny}
This form is exactly the same as that of Equation (\ref{product_rep}) so, if we apply the inverse of the Fourier transform, we  obtain $\ket{\varphi_1,\ldots,\varphi_t}$. The inverse of an operator is its adjoint so the overall phase estimation circuit is shown in Figure \ref{phase_est_all}.
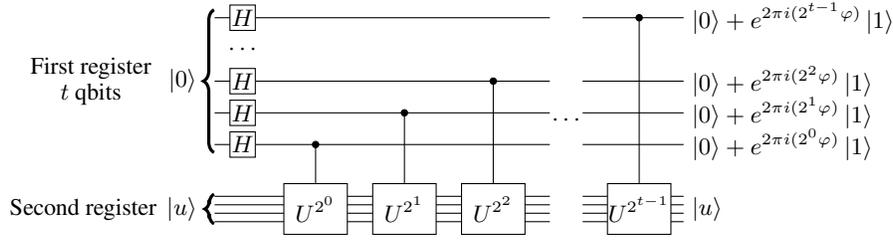
\begin{figure*}[t]
\centering
\begin{footnotesize}
\begin{tikzpicture}[scale=0.800000,x=1pt,y=1pt]
\filldraw[color=white] (0.000000, -2.000000) rectangle (222.000000, 104.000000);
\draw[color=black] (0.000000,96.500000) -- (222.000000,96.500000);
\draw[color=black] (0.000000,66.500000) -- (222.000000,66.500000);
\draw[color=black] (0.000000,51.500000) -- (222.000000,51.500000);
\draw[color=black] (0.000000,36.500000) -- (222.000000,36.500000);
\filldraw[color=white,fill=white] (0.000000,32.750000) rectangle (-4.000000,100.250000);
\draw[decorate,decoration={brace,amplitude = 4.000000pt},very thick] (0.000000,32.750000) -- (0.000000,100.250000);
\draw[color=black] (-4.000000,66.500000) node[left] {${\begin{array}{c}\mbox{First register}\\t\mbox{ qbits}\end{array}\ket{0}}$};
\draw[color=black] (0.000000,12.000000) -- (222.000000,12.000000);
\draw[color=black] (0.000000,8.000000) -- (222.000000,8.000000);
\draw[color=black] (0.000000,4.000000) -- (222.000000,4.000000);
\draw[color=black] (0.000000,0.000000) -- (222.000000,0.000000);
\filldraw[color=white,fill=white] (0.000000,-1.000000) rectangle (-4.000000,13.000000);
\draw[decorate,decoration={brace,amplitude = 3.500000pt},very thick] (0.000000,-1.000000) -- (0.000000,13.000000);
\draw[color=black] (-4.000000,6.000000) node[left] {${\mbox{Second register }\ket{u}}$};
\begin{scope}
\draw[fill=white] (13.500000, 96.500000) +(-45.000000:8.485281pt and 8.485281pt) -- +(45.000000:8.485281pt and 8.485281pt) -- +(135.000000:8.485281pt and 8.485281pt) -- +(225.000000:8.485281pt and 8.485281pt) -- cycle;
\clip (13.500000, 96.500000) +(-45.000000:8.485281pt and 8.485281pt) -- +(45.000000:8.485281pt and 8.485281pt) -- +(135.000000:8.485281pt and 8.485281pt) -- +(225.000000:8.485281pt and 8.485281pt) -- cycle;
\draw (13.500000, 96.500000) node {$H$};
\end{scope}
\begin{scope}
\draw[fill=white] (13.500000, 66.500000) +(-45.000000:8.485281pt and 8.485281pt) -- +(45.000000:8.485281pt and 8.485281pt) -- +(135.000000:8.485281pt and 8.485281pt) -- +(225.000000:8.485281pt and 8.485281pt) -- cycle;
\clip (13.500000, 66.500000) +(-45.000000:8.485281pt and 8.485281pt) -- +(45.000000:8.485281pt and 8.485281pt) -- +(135.000000:8.485281pt and 8.485281pt) -- +(225.000000:8.485281pt and 8.485281pt) -- cycle;
\draw (13.500000, 66.500000) node {$H$};
\end{scope}
\begin{scope}
\draw[fill=white] (13.500000, 51.500000) +(-45.000000:8.485281pt and 8.485281pt) -- +(45.000000:8.485281pt and 8.485281pt) -- +(135.000000:8.485281pt and 8.485281pt) -- +(225.000000:8.485281pt and 8.485281pt) -- cycle;
\clip (13.500000, 51.500000) +(-45.000000:8.485281pt and 8.485281pt) -- +(45.000000:8.485281pt and 8.485281pt) -- +(135.000000:8.485281pt and 8.485281pt) -- +(225.000000:8.485281pt and 8.485281pt) -- cycle;
\draw (13.500000, 51.500000) node {$H$};
\end{scope}
\begin{scope}
\draw[fill=white] (13.500000, 36.500000) +(-45.000000:8.485281pt and 8.485281pt) -- +(45.000000:8.485281pt and 8.485281pt) -- +(135.000000:8.485281pt and 8.485281pt) -- +(225.000000:8.485281pt and 8.485281pt) -- cycle;
\clip (13.500000, 36.500000) +(-45.000000:8.485281pt and 8.485281pt) -- +(45.000000:8.485281pt and 8.485281pt) -- +(135.000000:8.485281pt and 8.485281pt) -- +(225.000000:8.485281pt and 8.485281pt) -- cycle;
\draw (13.500000, 36.500000) node {$H$};
\end{scope}
\draw[fill=white,color=white] (6.000000, 75.500000) rectangle (21.000000, 87.500000);
\draw (13.500000, 81.500000) node {$\cdots$};
\draw (48.000000,36.500000) -- (48.000000,0.000000);
\begin{scope}
\draw[fill=white] (48.000000, 6.000000) +(-45.000000:21.213203pt and 16.970563pt) -- +(45.000000:21.213203pt and 16.970563pt) -- +(135.000000:21.213203pt and 16.970563pt) -- +(225.000000:21.213203pt and 16.970563pt) -- cycle;
\clip (48.000000, 6.000000) +(-45.000000:21.213203pt and 16.970563pt) -- +(45.000000:21.213203pt and 16.970563pt) -- +(135.000000:21.213203pt and 16.970563pt) -- +(225.000000:21.213203pt and 16.970563pt) -- cycle;
\draw (48.000000, 6.000000) node {$U^{2^0}$};
\end{scope}
\filldraw (48.000000, 36.500000) circle(1.500000pt);
\draw (90.000000,51.500000) -- (90.000000,0.000000);
\begin{scope}
\draw[fill=white] (90.000000, 6.000000) +(-45.000000:21.213203pt and 16.970563pt) -- +(45.000000:21.213203pt and 16.970563pt) -- +(135.000000:21.213203pt and 16.970563pt) -- +(225.000000:21.213203pt and 16.970563pt) -- cycle;
\clip (90.000000, 6.000000) +(-45.000000:21.213203pt and 16.970563pt) -- +(45.000000:21.213203pt and 16.970563pt) -- +(135.000000:21.213203pt and 16.970563pt) -- +(225.000000:21.213203pt and 16.970563pt) -- cycle;
\draw (90.000000, 6.000000) node {$U^{2^1}$};
\end{scope}
\filldraw (90.000000, 51.500000) circle(1.500000pt);
\draw (132.000000,66.500000) -- (132.000000,0.000000);
\begin{scope}
\draw[fill=white] (132.000000, 6.000000) +(-45.000000:21.213203pt and 16.970563pt) -- +(45.000000:21.213203pt and 16.970563pt) -- +(135.000000:21.213203pt and 16.970563pt) -- +(225.000000:21.213203pt and 16.970563pt) -- cycle;
\clip (132.000000, 6.000000) +(-45.000000:21.213203pt and 16.970563pt) -- +(45.000000:21.213203pt and 16.970563pt) -- +(135.000000:21.213203pt and 16.970563pt) -- +(225.000000:21.213203pt and 16.970563pt) -- cycle;
\draw (132.000000, 6.000000) node {$U^{2^2}$};
\end{scope}
\filldraw (132.000000, 66.500000) circle(1.500000pt);
\draw[fill=white,color=white] (159.000000, -6.000000) rectangle (174.000000, 102.500000);
\draw (166.500000, 48.250000) node {$\cdots$};
\draw (201.000000,96.500000) -- (201.000000,0.000000);
\begin{scope}
\draw[fill=white] (201.000000, 6.000000) +(-45.000000:21.213203pt and 16.970563pt) -- +(45.000000:21.213203pt and 16.970563pt) -- +(135.000000:21.213203pt and 16.970563pt) -- +(225.000000:21.213203pt and 16.970563pt) -- cycle;
\clip (201.000000, 6.000000) +(-45.000000:21.213203pt and 16.970563pt) -- +(45.000000:21.213203pt and 16.970563pt) -- +(135.000000:21.213203pt and 16.970563pt) -- +(225.000000:21.213203pt and 16.970563pt) -- cycle;
\draw (201.000000, 6.000000) node {$U^{2^{t-1}}$};
\end{scope}
\filldraw (201.000000, 96.500000) circle(1.500000pt);
\draw[color=black] (222.000000,96.500000) node[right] {${\ket{0}+e^{2\pi i(2^{t-1}\varphi)}\ket{1}}$};
\draw[color=black] (222.000000,66.500000) node[right] {${\ket{0}+e^{2\pi i(2^2\varphi)}\ket{1}}$};
\draw[color=black] (222.000000,51.500000) node[right] {${\ket{0}+e^{2\pi i(2^1\varphi)}\ket{1}}$};
\draw[color=black] (222.000000,36.500000) node[right] {${\ket{0}+e^{2\pi i(2^0\varphi)}\ket{1}}$};
\draw[color=black] (222.000000,6.000000) node[right] {${\ket{u}}$};
\end{tikzpicture}
\end{footnotesize}
\caption{First stage of phase estimation. On the right we have omitted normalization factors of $\frac{1}{\sqrt{2}}$.}
\label{phase_est}
\end{figure*}
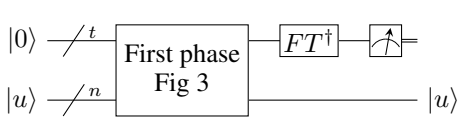
\begin{figure}[t]
\centering
\begin{tikzpicture}[scale=1.000000,x=1pt,y=1pt]
\filldraw[color=white] (0.000000, -7.500000) rectangle (140.000000, 37.500000);
\draw[color=black] (0.000000,22.500000) -- (128.000000,22.500000);
\draw[color=black] (128.000000,22.000000) -- (140.000000,22.000000);
\draw[color=black] (128.000000,23.000000) -- (140.000000,23.000000);
\draw[color=black] (0.000000,22.500000) node[left] {${\ket{0}}$};
\draw[color=black] (0.000000,0.000000) -- (140.000000,0.000000);
\draw[color=black] (0.000000,0.000000) node[left] {${\ket{u}}$};
\draw (6.000000, 16.500000) -- (14.000000, 28.500000);
\draw (12.000000, 25.500000) node[right] {$\scriptstyle{t}$};
\draw (6.000000, -6.000000) -- (14.000000, 6.000000);
\draw (12.000000, 3.000000) node[right] {$\scriptstyle{n}$};
\draw (51.000000,22.500000) -- (51.000000,0.000000);
\begin{scope}
\draw[fill=white] (51.000000, 11.250000) +(-45.000000:35.355339pt and 24.395184pt) -- +(45.000000:35.355339pt and 24.395184pt) -- +(135.000000:35.355339pt and 24.395184pt) -- +(225.000000:35.355339pt and 24.395184pt) -- cycle;
\clip (51.000000, 11.250000) +(-45.000000:35.355339pt and 24.395184pt) -- +(45.000000:35.355339pt and 24.395184pt) -- +(135.000000:35.355339pt and 24.395184pt) -- +(225.000000:35.355339pt and 24.395184pt) -- cycle;
\draw (51.000000, 11.250000) node {$\begin{array}{cc}\mbox{First phase}\\\mbox{Fig \ref{phase_est}}\end{array}$};
\end{scope}
\begin{scope}
\draw[fill=white] (99.000000, 22.500000) +(-45.000000:15.556349pt and 8.485281pt) -- +(45.000000:15.556349pt and 8.485281pt) -- +(135.000000:15.556349pt and 8.485281pt) -- +(225.000000:15.556349pt and 8.485281pt) -- cycle;
\clip (99.000000, 22.500000) +(-45.000000:15.556349pt and 8.485281pt) -- +(45.000000:15.556349pt and 8.485281pt) -- +(135.000000:15.556349pt and 8.485281pt) -- +(225.000000:15.556349pt and 8.485281pt) -- cycle;
\draw (99.000000, 22.500000) node {$FT^\dagger$};
\end{scope}
\draw[fill=white] (122.000000, 16.500000) rectangle (134.000000, 28.500000);
\draw[very thin] (128.000000, 23.100000) arc (90:150:6.000000pt);
\draw[very thin] (128.000000, 23.100000) arc (90:30:6.000000pt);
\draw[->,>=stealth] (128.000000, 17.100000) -- +(80:10.392305pt);
\draw[color=black] (140.000000,0.000000) node[right] {${\ket{u}}$};
\end{tikzpicture}
\caption{The complete phase estimation circuit.}
\label{phase_est_all}
\end{figure}

If $\varphi$ cannot be represented exactly with $t$ bits, the algorithm provides approximation guarantees: if we want to approximate $\varphi$ to $m$ bits with probability of success at least $1-\epsilon$ we must choose $t=m+\lceil \log_2 \left(2+\frac{1}{2\epsilon}\right)\rceil$  \cite{nielsen2010quantum}.

\section{Quantum Search}
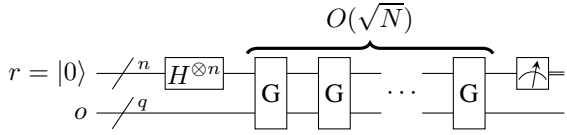
\begin{figure}[t]
\centering
\begin{tikzpicture}[scale=1.000000,x=1pt,y=1pt]
\filldraw[color=white] (0.000000, -7.500000) rectangle (177.000000, 22.500000);
\draw[color=black] (0.000000,15.000000) -- (165.000000,15.000000);
\draw[color=black] (165.000000,14.500000) -- (177.000000,14.500000);
\draw[color=black] (165.000000,15.500000) -- (177.000000,15.500000);
\draw[color=black] (0.000000,15.000000) node[left] {${r=\ket{0}}$};
\draw[color=black] (0.000000,0.000000) -- (177.000000,0.000000);
\draw[color=black] (0.000000,0.000000) node[left] {$o$};
\draw (6.000000, 9.000000) -- (14.000000, 21.000000);
\draw (12.000000, 18.000000) node[right] {$\scriptstyle{n}$};
\draw (6.000000, -6.000000) -- (14.000000, 6.000000);
\draw (12.000000, 3.000000) node[right] {$\scriptstyle{q}$};
\begin{scope}
\draw[fill=white] (37.000000, 15.000000) +(-45.000000:15.556349pt and 8.485281pt) -- +(45.000000:15.556349pt and 8.485281pt) -- +(135.000000:15.556349pt and 8.485281pt) -- +(225.000000:15.556349pt and 8.485281pt) -- cycle;
\clip (37.000000, 15.000000) +(-45.000000:15.556349pt and 8.485281pt) -- +(45.000000:15.556349pt and 8.485281pt) -- +(135.000000:15.556349pt and 8.485281pt) -- +(225.000000:15.556349pt and 8.485281pt) -- cycle;
\draw (37.000000, 15.000000) node {$H^{\otimes n}$};
\end{scope}
\draw (66.000000,15.000000) -- (66.000000,0.000000);
\begin{scope}
\draw[fill=white] (66.000000, 7.500000) +(-45.000000:8.485281pt and 19.091883pt) -- +(45.000000:8.485281pt and 19.091883pt) -- +(135.000000:8.485281pt and 19.091883pt) -- +(225.000000:8.485281pt and 19.091883pt) -- cycle;
\clip (66.000000, 7.500000) +(-45.000000:8.485281pt and 19.091883pt) -- +(45.000000:8.485281pt and 19.091883pt) -- +(135.000000:8.485281pt and 19.091883pt) -- +(225.000000:8.485281pt and 19.091883pt) -- cycle;
\draw (66.000000, 7.500000) node {G};
\end{scope}
\draw (90.000000,15.000000) -- (90.000000,0.000000);
\begin{scope}
\draw[fill=white] (90.000000, 7.500000) +(-45.000000:8.485281pt and 19.091883pt) -- +(45.000000:8.485281pt and 19.091883pt) -- +(135.000000:8.485281pt and 19.091883pt) -- +(225.000000:8.485281pt and 19.091883pt) -- cycle;
\clip (90.000000, 7.500000) +(-45.000000:8.485281pt and 19.091883pt) -- +(45.000000:8.485281pt and 19.091883pt) -- +(135.000000:8.485281pt and 19.091883pt) -- +(225.000000:8.485281pt and 19.091883pt) -- cycle;
\draw (90.000000, 7.500000) node {G};
\end{scope}
\draw[fill=white,color=white] (108.000000, -6.000000) rectangle (123.000000, 21.000000);
\draw (115.500000, 7.500000) node {$\cdots$};
\draw (141.000000,15.000000) -- (141.000000,0.000000);
\begin{scope}
\draw[fill=white] (141.000000, 7.500000) +(-45.000000:8.485281pt and 19.091883pt) -- +(45.000000:8.485281pt and 19.091883pt) -- +(135.000000:8.485281pt and 19.091883pt) -- +(225.000000:8.485281pt and 19.091883pt) -- cycle;
\clip (141.000000, 7.500000) +(-45.000000:8.485281pt and 19.091883pt) -- +(45.000000:8.485281pt and 19.091883pt) -- +(135.000000:8.485281pt and 19.091883pt) -- +(225.000000:8.485281pt and 19.091883pt) -- cycle;
\draw (141.000000, 7.500000) node {G};
\end{scope}
\draw[fill=white] (159.000000, 9.000000) rectangle (171.000000, 21.000000);
\draw[very thin] (165.000000, 15.600000) arc (90:150:6.000000pt);
\draw[very thin] (165.000000, 15.600000) arc (90:30:6.000000pt);
\draw[->,>=stealth] (165.000000, 9.600000) -- +(80:10.392305pt);
\draw[decorate,decoration={brace,amplitude = 4.000000pt},very thick] (57.000000,22.500000) -- (150.000000,22.500000);
\draw (103.500000, 26.500000) node[text width=144pt,above,text centered] {$O(\sqrt{N})$};
\end{tikzpicture}
\caption{Quantum search algorithm.}
\label{qs}
\end{figure}

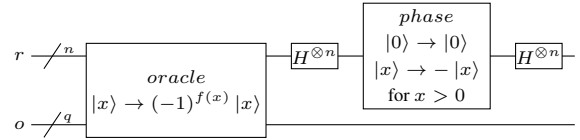
\begin{figure}[t]
\centering
\begin{scriptsize}
\begin{tikzpicture}[scale=0.800000,x=1pt,y=1pt]
\filldraw[color=white] (0.000000, -7.500000) rectangle (257.000000, 57.500000);
\draw[color=black] (0.000000,32.500000) -- (257.000000,32.500000);
\draw[color=black] (0.000000,32.500000) node[left] {$r$};
\draw[color=black] (0.000000,0.000000) -- (257.000000,0.000000);
\draw[color=black] (0.000000,0.000000) node[left] {$o$};
\draw (6.000000, 26.500000) -- (14.000000, 38.500000);
\draw (12.000000, 35.500000) node[right] {$\scriptstyle{n}$};
\draw (6.000000, -6.000000) -- (14.000000, 6.000000);
\draw (12.000000, 3.000000) node[right] {$\scriptstyle{q}$};
\draw (68.500000,32.500000) -- (68.500000,0.000000);
\begin{scope}
\draw[fill=white] (68.500000, 16.250000) +(-45.000000:60.104076pt and 31.466252pt) -- +(45.000000:60.104076pt and 31.466252pt) -- +(135.000000:60.104076pt and 31.466252pt) -- +(225.000000:60.104076pt and 31.466252pt) -- cycle;
\clip (68.500000, 16.250000) +(-45.000000:60.104076pt and 31.466252pt) -- +(45.000000:60.104076pt and 31.466252pt) -- +(135.000000:60.104076pt and 31.466252pt) -- +(225.000000:60.104076pt and 31.466252pt) -- cycle;
\draw (68.500000, 16.250000) node {$\begin{array}{c}oracle\\\ket{x}\to (-1)^{f(x)}\ket{x}\end{array}$};
\end{scope}
\begin{scope}
\draw[fill=white] (134.000000, 32.500000) +(-45.000000:15.556349pt and 8.485281pt) -- +(45.000000:15.556349pt and 8.485281pt) -- +(135.000000:15.556349pt and 8.485281pt) -- +(225.000000:15.556349pt and 8.485281pt) -- cycle;
\clip (134.000000, 32.500000) +(-45.000000:15.556349pt and 8.485281pt) -- +(45.000000:15.556349pt and 8.485281pt) -- +(135.000000:15.556349pt and 8.485281pt) -- +(225.000000:15.556349pt and 8.485281pt) -- cycle;
\draw (134.000000, 32.500000) node {$H^{\otimes n}$};
\end{scope}
\begin{scope}
\draw[fill=white] (187.000000, 32.500000) +(-45.000000:42.426407pt and 35.355339pt) -- +(45.000000:42.426407pt and 35.355339pt) -- +(135.000000:42.426407pt and 35.355339pt) -- +(225.000000:42.426407pt and 35.355339pt) -- cycle;
\clip (187.000000, 32.500000) +(-45.000000:42.426407pt and 35.355339pt) -- +(45.000000:42.426407pt and 35.355339pt) -- +(135.000000:42.426407pt and 35.355339pt) -- +(225.000000:42.426407pt and 35.355339pt) -- cycle;
\draw (187.000000, 32.500000) node {$\begin{array}{c}phase\\\ket{0}\to\ket{0}\\\ket{x}\to -\ket{x}\\\mbox{for }x>0\end{array}$};
\end{scope}
\begin{scope}
\draw[fill=white] (240.000000, 32.500000) +(-45.000000:15.556349pt and 8.485281pt) -- +(45.000000:15.556349pt and 8.485281pt) -- +(135.000000:15.556349pt and 8.485281pt) -- +(225.000000:15.556349pt and 8.485281pt) -- cycle;
\clip (240.000000, 32.500000) +(-45.000000:15.556349pt and 8.485281pt) -- +(45.000000:15.556349pt and 8.485281pt) -- +(135.000000:15.556349pt and 8.485281pt) -- +(225.000000:15.556349pt and 8.485281pt) -- cycle;
\draw (240.000000, 32.500000) node {$H^{\otimes n}$};
\end{scope}
\end{tikzpicture}
\end{scriptsize}
\caption{Grover operator.}
\label{grover}
\end{figure}

The problem of \emph{quantum search} is, given a Boolean function $\phi:\{0,1\}^n\to \{0,1\}$, return a configuration of bits $x$ such that $\phi(x)=1$ \cite{Grover:1996:FQM:237814.237866,grover1996fast,grover1997quantum}. We assume we have a black box that evaluates $\phi$, we call it an \emph{oracle} $O$, that is such that
$$\ket{x}\to^O(-1)^{\phi(x)}\ket{x}$$
i.e., the oracle marks solutions to the search problems by changing their sign. The oracle may use extra ancilla bits to do so. For the case of the function of Example \ref{sprinklerb}, the oracle will use a circuit such as the one of Figure \ref{sprinklerq} in its internals.
Figure \ref{qs} shows the circuit performing quantum search operating on an $n$-qubit register $r$ and the oracle workspace $o$. 
The circuit includes a gate $G$ that is called the \emph{Grover operator }and is implemented as show in Figure \ref{grover}.
The first gate of the search circuit applies the $H$ gate to each qubit in   register $r$ obtaining the 
\emph{uniform superposition} state
$$\ket{\psi}=\frac{1}{N^{1/2}}\sum_{x=0}^{N-1}\ket{x}$$
where $N=2^n$.

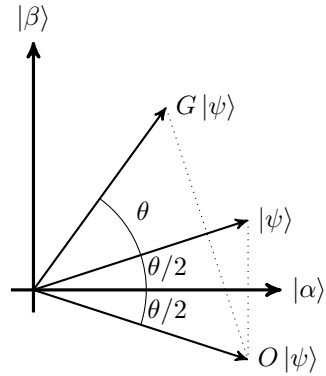
\begin{figure}[t]
\centering
\begin{tikzpicture}[
    scale=3,
    axis/.style={very thick, ->, >=stealth'},
    important line/.style={thick,->,>=stealth'},
    every node/.style={color=black}
    ]
    \draw[axis] (-0.1,0)  -- (1.1,0) coordinate(alpha) node(xline)[right]
        {$\ket{\alpha}$};
    \draw[axis] (0,-0.1) -- (0,1.1) node(yline)[above] {$\ket{\beta}$};
    \draw[important line] (0,0) coordinate (A) -- ({cos(\psiangle)},{sin(\psiangle)})
        coordinate (B) node[right, text width=5em] {$\ket{\psi}$};
    \draw[important line] (0,0) coordinate (C) -- ({cos(-\psiangle},{sin(-\psiangle)})
        coordinate (D) node[right, text width=5em] {$O\ket{\psi}$};
    \draw[important line] (0,0) coordinate (E) -- ({cos(3*\psiangle)},{sin(3*\psiangle)})
        coordinate (F) node[right, text width=5em] {$G\ket{\psi}$};
    \draw [dotted] (B) -- (D);
    \draw [dotted] (F) -- (D);
    \draw
    pic["$\theta/2$", draw=black, -, angle eccentricity=1.2, angle radius=1.5cm]
    {angle=D--A--alpha};
    \draw
    pic["$\theta/2$", draw=black, -, angle eccentricity=1.2, angle radius=1.5cm]
    {angle=alpha--A--B};
    \draw
    pic["$\theta$", draw=black, -, angle eccentricity=1.2, angle radius=1.5cm]
    {angle=B--A--F};
\end{tikzpicture}
\caption{Visualization of the effect of Grover operator.}
\label{plane}
\end{figure}

The Grover operator can be written as 
$$G=(2\ket{\psi}\bra{\psi}-I)O$$
We now show that the Grover operator is a rotation.
Consider the two states
$$\ket{\alpha}=\frac{1}{\sqrt{N-M}}\sum_{x:\phi(x)=0}\ket{x}$$
$$\ket{\beta}=\frac{1}{\sqrt{M}}\sum_{x:\phi(x)=1}\ket{x}$$
where $M$ is the number of solutions to $\phi(x)=1$. These two states are orthonormal.
The uniform superposition state $\ket{\psi}$ can be written as a linear combination of $\ket{\alpha}$ and
$\ket{\beta}$:
$$\ket{\psi}=\sqrt{\frac{N-M}{N}}\ket{\alpha}+\sqrt{\frac{M}{N}}\ket{\beta}$$
so $\ket{\psi}$ belongs to plane defined by $\ket{\alpha}$ and $\ket{\beta}$. In this plane, the effect of
the oracle
operation $O$ is to perform a reflection about the vector $\alpha$ because $O(\ket{\alpha} + \ket{\beta}) = \ket{\alpha} - \ket{\beta}$, see Figure \ref{plane}.

The other component of Grover operator, $2\ket{\psi}\bra{\psi}-I$, also performs a
reflection in the plane defined by $\ket{\alpha}$ and $\ket{\beta}$, about the vector $\ket{\psi}$.
The overall effect is that of a rotation \cite{aharonov1999quantum}. Define $\cos \theta/2=\sqrt{(N-M)/N}$, then $\ket{\psi}=\cos \theta/2\ket{\alpha}+\sin\theta/2\ket{\beta}$.


From Figure \ref{plane} we can see that the rotation applied by $G$ is exactly $\theta$ so 
$$G\ket{\psi}=\cos \frac{3\theta}{2}\ket{\alpha}+\sin\frac{3\theta}{2}\ket{\beta}$$
Repeated applications of $G$ take the state
to
$$G^k\ket{\psi}=\cos \left(\frac{2k+1}{2}\theta\right)\ket{\alpha}+\sin\left(\frac{2k+1}{2}\theta\right)\ket{\beta}.$$
These rotations bring $\ket{\psi}$ closer and closer to $\ket{\beta}$. If we perform the right number of rotations, an observation in the
computational basis produces with high probability one of the outcomes superposed in
$\ket{\beta}$, i.e., a solution to the search problem. It turns out that the number of applications of $G$ (and thus of oracle calls) required to maximise the probability of measuring one of the solutions to the search problem is $O(\sqrt{N/M})$, while classically by treating $\phi$ as a black box the number of oracle calls would be $O(N/M)$.

The algorithm works if $M\leq N/2$. If this is not true, it is enough to consider an extra qubit $e$, defining a new function $\phi'(x)$ that is true only if $e$ is true, i.e., $\phi'(x)=\phi(x)\wedge e$. This leaves $M$ unchanged but multiplies $N$ by 2.

\section{Quantum Counting}
With quantum counting we want to count the number of solutions to the equation $\phi(x)=1$ where $\phi$ is a Boolean function as above. In the notation of the previous section, it means computing $M$.

Suppose $\ket{a}$ and $\ket{b}$ are the two eigenvectors of  the Grover operator $G$ in the space spanned by $\ket{\alpha}$ and $\ket{\beta}$. Since $G$ is a rotation of angle $\theta$ in such a space, the eigenvalues of $\ket{a}$ and $\ket{b}$ are $e^{i\theta}$ and $e^{i(2\pi-\theta)}$.
If we know $\theta$, we can compute $M$ from $\sin^2(\theta/2) = M/2N$ (supposing the oracle has been augmented). Since $\sin(\theta/2)=\sin(\pi-\theta/2)$, it does not matter which eigenvalue is estimated.

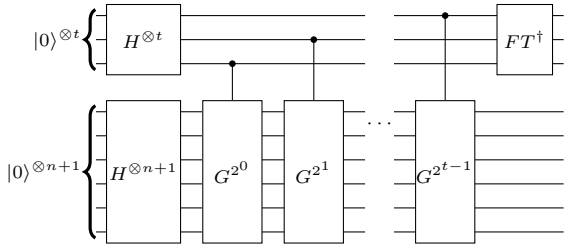
\begin{figure}[t]
\centering
\begin{scriptsize}
\begin{tikzpicture}[scale=0.700000,x=1pt,y=1pt]
\filldraw[color=white] (0.000000, -6.500000) rectangle (253.000000, 123.500000);
\draw[color=black] (0.000000,117.000000) -- (253.000000,117.000000);
\draw[color=black] (0.000000,104.000000) -- (253.000000,104.000000);
\draw[color=black] (0.000000,91.000000) -- (253.000000,91.000000);
\filldraw[color=white,fill=white] (0.000000,87.750000) rectangle (-4.000000,120.250000);
\draw[decorate,decoration={brace,amplitude = 4.000000pt},very thick] (0.000000,87.750000) -- (0.000000,120.250000);
\draw[color=black] (-4.000000,104.000000) node[left] {${\ket{0}^{\otimes t}}$};
\draw[color=black] (0.000000,65.000000) -- (253.000000,65.000000);
\draw[color=black] (0.000000,52.000000) -- (253.000000,52.000000);
\draw[color=black] (0.000000,39.000000) -- (253.000000,39.000000);
\draw[color=black] (0.000000,26.000000) -- (253.000000,26.000000);
\draw[color=black] (0.000000,13.000000) -- (253.000000,13.000000);
\draw[color=black] (0.000000,0.000000) -- (253.000000,0.000000);
\filldraw[color=white,fill=white] (0.000000,-3.250000) rectangle (-4.000000,68.250000);
\draw[decorate,decoration={brace,amplitude = 4.000000pt},very thick] (0.000000,-3.250000) -- (0.000000,68.250000);
\draw[color=black] (-4.000000,32.500000) node[left] {${\ket{0}^{\otimes n+1}}$};
\draw (26.000000,117.000000) -- (26.000000,91.000000);
\begin{scope}
\draw[fill=white] (26.000000, 104.000000) +(-45.000000:28.284271pt and 26.870058pt) -- +(45.000000:28.284271pt and 26.870058pt) -- +(135.000000:28.284271pt and 26.870058pt) -- +(225.000000:28.284271pt and 26.870058pt) -- cycle;
\clip (26.000000, 104.000000) +(-45.000000:28.284271pt and 26.870058pt) -- +(45.000000:28.284271pt and 26.870058pt) -- +(135.000000:28.284271pt and 26.870058pt) -- +(225.000000:28.284271pt and 26.870058pt) -- cycle;
\draw (26.000000, 104.000000) node {$H^{\otimes t}$};
\end{scope}
\draw (26.000000,65.000000) -- (26.000000,0.000000);
\begin{scope}
\draw[fill=white] (26.000000, 32.500000) +(-45.000000:28.284271pt and 54.447222pt) -- +(45.000000:28.284271pt and 54.447222pt) -- +(135.000000:28.284271pt and 54.447222pt) -- +(225.000000:28.284271pt and 54.447222pt) -- cycle;
\clip (26.000000, 32.500000) +(-45.000000:28.284271pt and 54.447222pt) -- +(45.000000:28.284271pt and 54.447222pt) -- +(135.000000:28.284271pt and 54.447222pt) -- +(225.000000:28.284271pt and 54.447222pt) -- cycle;
\draw (26.000000, 32.500000) node {$H^{\otimes n+1}$};
\end{scope}
\draw (74.000000,91.000000) -- (74.000000,0.000000);
\begin{scope}
\draw[fill=white] (74.000000, 32.500000) +(-45.000000:22.627417pt and 54.447222pt) -- +(45.000000:22.627417pt and 54.447222pt) -- +(135.000000:22.627417pt and 54.447222pt) -- +(225.000000:22.627417pt and 54.447222pt) -- cycle;
\clip (74.000000, 32.500000) +(-45.000000:22.627417pt and 54.447222pt) -- +(45.000000:22.627417pt and 54.447222pt) -- +(135.000000:22.627417pt and 54.447222pt) -- +(225.000000:22.627417pt and 54.447222pt) -- cycle;
\draw (74.000000, 32.500000) node {$G^{2^0}$};
\end{scope}
\filldraw (74.000000, 91.000000) circle(1.500000pt);
\draw (118.000000,104.000000) -- (118.000000,0.000000);
\begin{scope}
\draw[fill=white] (118.000000, 32.500000) +(-45.000000:22.627417pt and 54.447222pt) -- +(45.000000:22.627417pt and 54.447222pt) -- +(135.000000:22.627417pt and 54.447222pt) -- +(225.000000:22.627417pt and 54.447222pt) -- cycle;
\clip (118.000000, 32.500000) +(-45.000000:22.627417pt and 54.447222pt) -- +(45.000000:22.627417pt and 54.447222pt) -- +(135.000000:22.627417pt and 54.447222pt) -- +(225.000000:22.627417pt and 54.447222pt) -- cycle;
\draw (118.000000, 32.500000) node {$G^{2^1}$};
\end{scope}
\filldraw (118.000000, 104.000000) circle(1.500000pt);
\draw[fill=white,color=white] (146.000000, -6.000000) rectangle (161.000000, 123.000000);
\draw (153.500000, 58.500000) node {$\cdots$};
\draw (189.000000,117.000000) -- (189.000000,0.000000);
\begin{scope}
\draw[fill=white] (189.000000, 32.500000) +(-45.000000:22.627417pt and 54.447222pt) -- +(45.000000:22.627417pt and 54.447222pt) -- +(135.000000:22.627417pt and 54.447222pt) -- +(225.000000:22.627417pt and 54.447222pt) -- cycle;
\clip (189.000000, 32.500000) +(-45.000000:22.627417pt and 54.447222pt) -- +(45.000000:22.627417pt and 54.447222pt) -- +(135.000000:22.627417pt and 54.447222pt) -- +(225.000000:22.627417pt and 54.447222pt) -- cycle;
\draw (189.000000, 32.500000) node {$G^{2^{t-1}}$};
\end{scope}
\filldraw (189.000000, 117.000000) circle(1.500000pt);
\draw (232.000000,117.000000) -- (232.000000,91.000000);
\begin{scope}
\draw[fill=white] (232.000000, 104.000000) +(-45.000000:21.213203pt and 26.870058pt) -- +(45.000000:21.213203pt and 26.870058pt) -- +(135.000000:21.213203pt and 26.870058pt) -- +(225.000000:21.213203pt and 26.870058pt) -- cycle;
\clip (232.000000, 104.000000) +(-45.000000:21.213203pt and 26.870058pt) -- +(45.000000:21.213203pt and 26.870058pt) -- +(135.000000:21.213203pt and 26.870058pt) -- +(225.000000:21.213203pt and 26.870058pt) -- cycle;
\draw (232.000000, 104.000000) node {$FT^\dagger$};
\end{scope}
\end{tikzpicture}
\end{scriptsize}
\caption{Circuit for quantum counting.}
\label{quantumcounting}
\end{figure}

So quantum counting is performed by using quantum phase estimation to compute the eigenvalues of the 
Grover operator $G$. The circuit for quantum counting is shown in Figure \ref{quantumcounting} \cite{boyer1998tight,DBLP:conf/icalp/BrassardHT98}.

The upper register in Figure \ref{quantumcounting} has $t$ qubits while the lower register $n+1$. 
$\theta$ is estimated to $m$ bits of accuray with probability at least $1-\epsilon$ if $t=m+\lceil \log_2 (2+1/2\epsilon)\rceil$. The error on the estimate of the count $M$ is given by  
\cite{nielsen2010quantum}:
\begin{scriptsize}
\begin{eqnarray*}
&&\frac{|\Delta M|}{2N}=\left|\sin^2\left(\frac{\theta+\Delta\theta}{2}\right)-\sin^2\left(\frac{\theta}{2}\right)\right|=\\
&&\left(\sin\left(\frac{\theta+\Delta\theta}{2}\right)+\sin\left(\frac{\theta}{2}\right)\right)\left|\sin\left(\frac{\theta+\Delta\theta}{2}\right)-\sin\left(\frac{\theta}{2}\right)\right|
\end{eqnarray*}
\end{scriptsize}
Since $|\sin((\theta+\Delta\theta)/2)-\sin(\theta/2)|\leq|\Delta\theta|/2$ and $|\sin((\theta+\Delta\theta)/2)|<\sin(\theta/2)+|\Delta\theta|/2$ from calculus and trigonometry respectively, we get
$$\frac{|\Delta M|}{2N}<\left(2\sin\left(\frac{\theta}{2}\right)+\frac{|\Delta\theta|}{2}\right)\frac{|\Delta\theta|}{2}$$
Using $\sin^2(\theta/2)=M/2N$ and $|\Delta\theta|\leq 2^{-m}$ we obtain
$$|\Delta M|<\left(\sqrt{2MN}+\frac{N}{2^{m+1}}\right)2^{-m}$$
Consider this case: let $m=\lceil n/2\rceil+2$ and $\epsilon=1/12$. Then $t=\lceil n/2\rceil+5$. The number of applications of the Grover operator is $\Theta(\sqrt{N})$ and so is the number of oracle calls. The error is  $|\Delta M|<\sqrt{M/8}+1/32=O(\sqrt{M})$.

\section{Quantum Weighted Model Counting}

For the moment suppose that the literal weights sum to 1, i.e., that $w(x_i)+w(\neg x_i)=1$ for all
bits $x_i$.

\begin{figure}[t]
\centering
\begin{scriptsize}
\begin{tikzpicture}[scale=0.800000,x=1pt,y=1pt]
\filldraw[color=white] (0.000000, -3.500000) rectangle (237.000000, 143.500000);
\draw[color=black] (0.000000,140.000000) -- (237.000000,140.000000);
\draw[color=black] (0.000000,133.000000) -- (237.000000,133.000000);
\draw[color=black] (0.000000,126.000000) -- (237.000000,126.000000);
\filldraw[color=white,fill=white] (0.000000,124.250000) rectangle (-4.000000,141.750000);
\draw[decorate,decoration={brace,amplitude = 4.000000pt},very thick] (0.000000,124.250000) -- (0.000000,141.750000);
\draw[color=black] (-4.000000,133.000000) node[left] {${\begin{array}{c}\mbox{Register 1}\\\ket{0}^{\otimes t}\end{array}}$};
\draw[color=black] (0.000000,100.000000) -- (237.000000,100.000000);
\draw[color=black] (0.000000,85.000000) -- (237.000000,85.000000);
\draw[color=black] (0.000000,70.000000) -- (237.000000,70.000000);
\draw[color=black] (0.000000,40.000000) -- (237.000000,40.000000);
\draw[color=black] (0.000000,25.000000) -- (237.000000,25.000000);
\filldraw[color=white,fill=white] (0.000000,21.250000) rectangle (-4.000000,103.750000);
\draw[decorate,decoration={brace,amplitude = 4.000000pt},very thick] (0.000000,21.250000) -- (0.000000,103.750000);
\draw[color=black] (-4.000000,62.500000) node[left] {${\begin{array}{c}\mbox{Register 2}\\\ket{0}^{\otimes n+1}\end{array}}$};
\draw[color=black] (0.000000,14.000000) -- (237.000000,14.000000);
\draw[color=black] (0.000000,7.000000) -- (237.000000,7.000000);
\draw[color=black] (0.000000,0.000000) -- (237.000000,0.000000);
\filldraw[color=white,fill=white] (0.000000,-1.750000) rectangle (-4.000000,15.750000);
\draw[decorate,decoration={brace,amplitude = 4.000000pt},very thick] (0.000000,-1.750000) -- (0.000000,15.750000);
\draw[color=black] (-4.000000,7.000000) node[left] {${\begin{array}{c}\mbox{Ancilla}\\\ket{0}^{\otimes q}\end{array}}$};
\draw (21.000000,140.000000) -- (21.000000,126.000000);
\begin{scope}
\draw[fill=white] (21.000000, 133.000000) +(-45.000000:21.213203pt and 18.384776pt) -- +(45.000000:21.213203pt and 18.384776pt) -- +(135.000000:21.213203pt and 18.384776pt) -- +(225.000000:21.213203pt and 18.384776pt) -- cycle;
\clip (21.000000, 133.000000) +(-45.000000:21.213203pt and 18.384776pt) -- +(45.000000:21.213203pt and 18.384776pt) -- +(135.000000:21.213203pt and 18.384776pt) -- +(225.000000:21.213203pt and 18.384776pt) -- cycle;
\draw (21.000000, 133.000000) node {$H^{\otimes t}$};
\end{scope}
\begin{scope}
\draw[fill=white] (21.000000, 100.000000) +(-45.000000:21.213203pt and 8.485281pt) -- +(45.000000:21.213203pt and 8.485281pt) -- +(135.000000:21.213203pt and 8.485281pt) -- +(225.000000:21.213203pt and 8.485281pt) -- cycle;
\clip (21.000000, 100.000000) +(-45.000000:21.213203pt and 8.485281pt) -- +(45.000000:21.213203pt and 8.485281pt) -- +(135.000000:21.213203pt and 8.485281pt) -- +(225.000000:21.213203pt and 8.485281pt) -- cycle;
\draw (21.000000, 100.000000) node {$R_y(\theta_1)$};
\end{scope}
\begin{scope}
\draw[fill=white] (21.000000, 85.000000) +(-45.000000:21.213203pt and 8.485281pt) -- +(45.000000:21.213203pt and 8.485281pt) -- +(135.000000:21.213203pt and 8.485281pt) -- +(225.000000:21.213203pt and 8.485281pt) -- cycle;
\clip (21.000000, 85.000000) +(-45.000000:21.213203pt and 8.485281pt) -- +(45.000000:21.213203pt and 8.485281pt) -- +(135.000000:21.213203pt and 8.485281pt) -- +(225.000000:21.213203pt and 8.485281pt) -- cycle;
\draw (21.000000, 85.000000) node {$R_y(\theta_2)$};
\end{scope}
\begin{scope}
\draw[fill=white] (21.000000, 70.000000) +(-45.000000:21.213203pt and 8.485281pt) -- +(45.000000:21.213203pt and 8.485281pt) -- +(135.000000:21.213203pt and 8.485281pt) -- +(225.000000:21.213203pt and 8.485281pt) -- cycle;
\clip (21.000000, 70.000000) +(-45.000000:21.213203pt and 8.485281pt) -- +(45.000000:21.213203pt and 8.485281pt) -- +(135.000000:21.213203pt and 8.485281pt) -- +(225.000000:21.213203pt and 8.485281pt) -- cycle;
\draw (21.000000, 70.000000) node {$R_y(\theta_3)$};
\end{scope}
\begin{scope}
\draw (21.000000, 55.000000) node {$\cdots$};
\end{scope}
\begin{scope}
\draw[fill=white] (21.000000, 40.000000) +(-45.000000:21.213203pt and 8.485281pt) -- +(45.000000:21.213203pt and 8.485281pt) -- +(135.000000:21.213203pt and 8.485281pt) -- +(225.000000:21.213203pt and 8.485281pt) -- cycle;
\clip (21.000000, 40.000000) +(-45.000000:21.213203pt and 8.485281pt) -- +(45.000000:21.213203pt and 8.485281pt) -- +(135.000000:21.213203pt and 8.485281pt) -- +(225.000000:21.213203pt and 8.485281pt) -- cycle;
\draw (21.000000, 40.000000) node {$R_y(\theta_n)$};
\end{scope}
\begin{scope}
\draw[fill=white] (21.000000, 25.000000) +(-45.000000:8.485281pt and 8.485281pt) -- +(45.000000:8.485281pt and 8.485281pt) -- +(135.000000:8.485281pt and 8.485281pt) -- +(225.000000:8.485281pt and 8.485281pt) -- cycle;
\clip (21.000000, 25.000000) +(-45.000000:8.485281pt and 8.485281pt) -- +(45.000000:8.485281pt and 8.485281pt) -- +(135.000000:8.485281pt and 8.485281pt) -- +(225.000000:8.485281pt and 8.485281pt) -- cycle;
\draw (21.000000, 25.000000) node {$H$};
\end{scope}
\draw (63.000000,126.000000) -- (63.000000,0.000000);
\begin{scope}
\draw[fill=white] (63.000000, 50.000000) +(-45.000000:21.213203pt and 79.195959pt) -- +(45.000000:21.213203pt and 79.195959pt) -- +(135.000000:21.213203pt and 79.195959pt) -- +(225.000000:21.213203pt and 79.195959pt) -- cycle;
\clip (63.000000, 50.000000) +(-45.000000:21.213203pt and 79.195959pt) -- +(45.000000:21.213203pt and 79.195959pt) -- +(135.000000:21.213203pt and 79.195959pt) -- +(225.000000:21.213203pt and 79.195959pt) -- cycle;
\draw (63.000000, 50.000000) node {$G^{2^0}$};
\end{scope}
\filldraw (63.000000, 126.000000) circle(1.500000pt);
\draw (105.000000,133.000000) -- (105.000000,0.000000);
\begin{scope}
\draw[fill=white] (105.000000, 50.000000) +(-45.000000:21.213203pt and 79.195959pt) -- +(45.000000:21.213203pt and 79.195959pt) -- +(135.000000:21.213203pt and 79.195959pt) -- +(225.000000:21.213203pt and 79.195959pt) -- cycle;
\clip (105.000000, 50.000000) +(-45.000000:21.213203pt and 79.195959pt) -- +(45.000000:21.213203pt and 79.195959pt) -- +(135.000000:21.213203pt and 79.195959pt) -- +(225.000000:21.213203pt and 79.195959pt) -- cycle;
\draw (105.000000, 50.000000) node {$G^{2^1}$};
\end{scope}
\filldraw (105.000000, 133.000000) circle(1.500000pt);
\draw[fill=white,color=white] (132.000000, -6.000000) rectangle (147.000000, 146.000000);
\draw (139.500000, 70.000000) node {$\cdots$};
\draw (174.000000,140.000000) -- (174.000000,0.000000);
\begin{scope}
\draw[fill=white] (174.000000, 50.000000) +(-45.000000:21.213203pt and 79.195959pt) -- +(45.000000:21.213203pt and 79.195959pt) -- +(135.000000:21.213203pt and 79.195959pt) -- +(225.000000:21.213203pt and 79.195959pt) -- cycle;
\clip (174.000000, 50.000000) +(-45.000000:21.213203pt and 79.195959pt) -- +(45.000000:21.213203pt and 79.195959pt) -- +(135.000000:21.213203pt and 79.195959pt) -- +(225.000000:21.213203pt and 79.195959pt) -- cycle;
\draw (174.000000, 50.000000) node {$G^{2^{t-1}}$};
\end{scope}
\filldraw (174.000000, 140.000000) circle(1.500000pt);
\draw (216.000000,140.000000) -- (216.000000,126.000000);
\begin{scope}
\draw[fill=white] (216.000000, 133.000000) +(-45.000000:21.213203pt and 18.384776pt) -- +(45.000000:21.213203pt and 18.384776pt) -- +(135.000000:21.213203pt and 18.384776pt) -- +(225.000000:21.213203pt and 18.384776pt) -- cycle;
\clip (216.000000, 133.000000) +(-45.000000:21.213203pt and 18.384776pt) -- +(45.000000:21.213203pt and 18.384776pt) -- +(135.000000:21.213203pt and 18.384776pt) -- +(225.000000:21.213203pt and 18.384776pt) -- cycle;
\draw (216.000000, 133.000000) node {$FT^\dagger$};
\end{scope}
\end{tikzpicture}
\end{scriptsize}
\caption{Circuit for quantum weighted model counting.}
\label{qwmc_fig}
\end{figure}
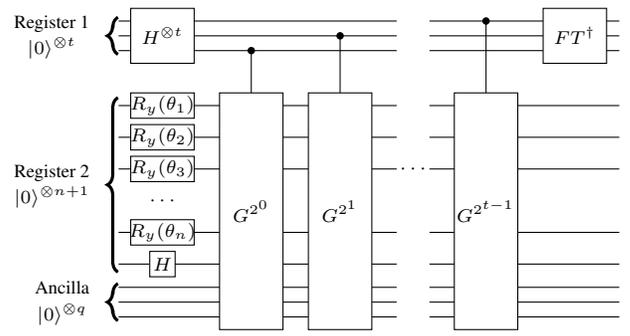

The circuit for performing quantum weighted model counting is shown in Figure \ref{qwmc_fig} and differs from the one in Figure \ref{quantumcounting} because the Hadamard operations applied to the lower register are replaced
by rotations $R_y(\theta_i)$ where $i$ is the qubit index except for the extra qubit for which the Hadamard operator is kept.
$\theta_i$ is computed as
$$\theta_i=2\arccos \sqrt{1-w_i}$$
where $w_i=w(x_i)$.
So
$$\cos \theta_i/2=\cos \arccos \sqrt{1-w_i}=\sqrt{1-w_i}$$
and
$$\sin \theta_i/2=\sqrt{1-(\cos\theta_i/2)^2}=\sqrt{w_i}$$
The effect of the rotation on the $i$th bit is
$$R_y(\theta_i)\ket{0}=\left[\begin{array}{cc}
\cos\frac{\theta_i}{2}&-\sin\frac{\theta_i}{2}\\
\sin\frac{\theta_i}{2}&\cos\frac{\theta_i}{2}
\end{array}\right] \left[\begin{array}{c}
1\\
0
\end{array}
\right]=\left[\begin{array}{c}
\cos\frac{\theta_i}{2}\\
\sin\frac{\theta_i}{2}\
\end{array}
\right]
=$$
$$\left[\begin{array}{c}
\sqrt{1-w_i}\\
\sqrt{w_i}
\end{array}
\right]=\sqrt{1-w_i}\ket{0}+\sqrt{w_i}\ket{1}
$$
Therefore the rotations prepare the state
$$\psi=\bigotimes_{i=1}^n( \sqrt{1-w_i}\ket{0}+\sqrt{w_i}\ket{1})\otimes \frac{1}{\sqrt{2}}(\ket{0}+\ket{1})=$$

$$=\sum_{b_{n+1}b_{n}\ldots b_1=0}^{2^{n+1}-1}\sqrt{0.5w'_{{n}}\ldots w'_{1}}\ket{b_{n+1}b_{n}\ldots b_1}$$
where $w'_{i}$ is
$$w'_{i}=\left\{\begin{array}{ll}
w_i&\mbox{if }b_i=1\\
1-w_i&\mbox{if }b_i=0
\end{array}\right.
$$
Define $W_{b_{n}b_{n-1}\ldots b_1}$ as $w'_{{n}}w'_{{n-1}}\ldots w'_{1}$
and normalized states
$$\ket{\alpha}=\frac{1}{\sqrt{0.5\sum_{x;\phi(x)=0}W_x}}\sum_{x;\phi(x)=0}\sqrt{0.5W_x}\ket{x}$$
$$\ket{\beta}=\frac{1}{\sqrt{0.5\sum_{x;\phi(x)=1}W_x}}\sum_{x;\phi(x)=1}\sqrt{0.5W_x}\ket{x},$$
then $\ket{\psi}$ can be expressed as
$$\ket{\psi}=\left(\sqrt{0.5\sum_{x;\phi(x)=0}W_x}\right)\ket{\alpha}+\left(\sqrt{0.5\sum_{x;\phi(x)=1}W_x}\right)\ket{\beta}$$
so the initial state of the quantum computer is in the space spanned by $\ket{\alpha}$ and $\ket{\beta}$

Let $\cos \theta/2=\sqrt{0.5\sum_{x;\phi(x)=0}W_x}$ and
 $\sin \theta/2=\sqrt{0.5\sum_{x;\phi(x)=1}W_x}$
 so that 
$$\ket{\psi}=\cos\theta/2\ket{\alpha}+\sin\theta/2\ket{\beta}$$
From this point we can repeat the reasoning used for quantum counting: the application of the 
Grover operator rotates $\ket{\psi}$ in the space spanned by $\ket{\alpha}$ and $\ket{\beta}$ by angle
 $\theta$ and  $e^{i\theta}$ and $e^{i(2\pi-\theta)}$ are the eigenvalues of $G$. 
$\theta$ can be found by quantum phase estimation.
From $\sin^2(\theta/2) =0.5\sum_{x;\phi(x)=1}W_x$ we obtain
$$WMC(\phi,w)=\sum_{x:\phi(x)=1}W_x=2\sin^2(\theta/2)$$
If the literal weights do not sum to 1, i.e., $w(x_i)+w(\neg x_i)\neq 1$, consider the normalized weights, i.e., the new weights
$\hat w(x_i)=\frac{w(x_i)}{w(x_i)+w(\neg x_i)}$ and $\hat w (\neg x_i)=\frac{w(\neg x_i)}{w(x_i)+w(\neg x_i)}$. Let $V_i$ be $w(x_i)+w(\neg x_i)$ for $i=1,\ldots,n$.
Then we perform QWMC with $\hat w$ replacing $w$. We get a normalized WMC $\widehat{WMC}(\phi,w)$
$$\widehat{WMC}(\phi,w)=\sum_{x:\phi(x)=1}\hat W_x$$ 
where $\hat W_{b_{n}b_{n-1}\ldots b_1}$ is $\hat w'_{{n}}\hat w'_{{n-1}}\ldots \hat w'_{1}$
and
$$\hat w'_{i}=\left\{\begin{array}{ll}
\hat w(x_i)&\mbox{if }b_i=1\\
1-\hat w(x_i)&\mbox{if }b_i=0
\end{array}\right.
$$
Then
\begin{eqnarray*}
&&\widehat{WMC}(\phi,w)=\\
&&
\sum_{x:\phi(x)=1}\hat W_x=\\
&&\sum_{b_{n}\ldots b_1:\phi(b_{n}\ldots b_1)=1}\hat W_{b_{n}\ldots b_1}=\\
&&\sum_{b_{n}\ldots b_1:\phi(b_{n}\ldots b_1)=1}\hat w'_{{n}}\ldots \hat w'_{1}=\\
&&\sum_{b_{n}\ldots b_1:\phi(b_{n}\ldots b_1)=1}\frac{w(b_{n})}{V_{n}}\ldots \frac{w(b_1)}{V_1}=\\
&&\sum_{b_{n}\ldots b_1:\phi(b_{n}\ldots b_1)=1}\frac{1}{\prod_{i=1}^{n}V_i} w(b_{n}) \ldots  w(b_1)=\\
&&\frac{1}{\prod_{i=1}^{n}V_i}\sum_{b_{n}\ldots b_1:\phi(b_{n}\ldots b_1)=1} w(b_{n}) \ldots  w(b_1)=\\
&&\frac{1}{\prod_{i=1}^{n}V_i}WMC(\phi,w)
\end{eqnarray*}
where $w(b_i)=w(x_i)$ if $b_i=1$ and $w(b_i)=w(\neg x_i)$ if $b_i=0$.
So if we multiply  $\widehat{WMC}(\phi,w)$ by ${\prod_{i=1}^{n}V_i}$ we obtain $WMC(\phi,w)$ also when the weights do not sum to 1.

Let us consider the complexity of the algorithm. We can repeat the derivation of the previous section where $M$ is replaced by  $N\times \widehat{WMC}(\phi,w)$.
We get
$$\frac{|\Delta \widehat{WMC}(\phi,w)|}{2}<\left(2\sin\left(\frac{\theta}{2}\right)+\frac{|\Delta\theta|}{2}\right)\frac{|\Delta\theta|}{2}$$
Using $\sin^2(\theta/2)=\widehat{WMC}(\phi,w)/2$ and $|\Delta\theta|\leq 2^{-m}$ we obtain
$$|\Delta \widehat{WMC}(\phi,w)|<\left(\sqrt{2\widehat{WMC}(\phi,w)}+2^{-m-1}\right)2^{-m}$$
Since $\widehat{WMC}(\phi,w)\leq1$ we have
\begin{small}
$$|\Delta \widehat{WMC}(\phi,w)|<\left(\sqrt{2}+2^{-m-1}\right)2^{-m}<2^{-m+\frac{1}{2}}+2^{-2m-1}
$$
\end{small}
If we choose $m=\lceil n/2\rceil+2$ and $\epsilon=1/12$, then $t=\lceil n/2\rceil+5$ and the algorithm 
requires $\Theta(\sqrt{N})$ oracle calls. The error becomes (for $n$ even, for $n$ odd the result is similar):
\begin{eqnarray*}
&&|\Delta \widehat{WMC}(\phi,w)|<2^{-\frac{n}{2}-2+\frac{1}{2}}+2^{-n-5}<\\
&&2^{-\frac{n}{2}-\frac{3}{2}}+2^{-\frac{n}{2}-\frac{3}{2}}<2^{-\frac{n}{2}-\frac{1}{2}}
\end{eqnarray*} 
so the error is bounded by $2^{-\frac{n+1}{2}}$.
%
%
%
%
%

\section{Complexity of Classical Algorithms}

Let us now discuss the advantages fo QWMC with respect to WMC. We consider a black box model of computation \cite{nielsen2010quantum}, where the only knowledge we have on the Boolean function $\phi$ is the possibility of evaluating it given an assignment of the Boolean variables, i.e., we have an oracle that answers queries over $\phi$. We want to know what is the minimum number 
of evaluations that are needed to solve counting problems.

Consider first an unweighted counting problem. A classical algorithm for probabilistically solving it proceeds by taking $k$ samples uniformly from the search space. This can be performed by sampling each Boolean variable uniformly and combining the bit samples obtaining an assignment sample. For each assignment sample, we query the oracle and we obtain a value $X_i$ with $i=1,\ldots,k$, where $X_i$ is 1 if $\phi$ evaluates to true for the sample and $X_i$ is 0 if $\phi$ evaluates to false. Then we can estimate the count as
$$S=\frac{N}{k}\times\sum_{i=1}^kX_k=\frac{N\overline{X}}{k}$$
where $\overline{X}=\sum_{i=1}^kX_k$.
Variable $\overline{X}=Sk/N$ is binomially distributed with $k$ the number of trials and probability of success $M/N$ where $M$ is the model count of $\phi$. Therefore the mean of $\overline{X}$ is $kM/N$ and the mean of $S$ is $N/kkM/N=M$, so $S$ is unbiased estimate of $M$. 

If we want to have probability at least 3/4 of estimating $M$ within an accuracy of $\sqrt{M}$ we can use the normal approximation of the binomial proportion confidence interval according to which the 
true success probability of the binomial variable lies in the interval
$$\hat p \pm z \sqrt{\frac{\hat p \left(1 - \hat p\right)}{k}}$$
where $\hat p$ is the estimated probability and  $z$ is the 
quantile
 of a standard normal distribution that depends on the confidence (in our case the confidence is 75\% and so $z=0.6744898$).
The size of the interval where the true probability lies is therefore 
$$2z\sqrt{\frac{S/N(1-S/N)}{k}}$$
and the the size of the interval of the number of solutions is
$$2zN\sqrt{\frac{S/N(1-S/N)}{k}}.$$
We replace the estimated probability with the true one to get a better estimate:
$$2zN\sqrt{\frac{M/N(1-M/N)}{k}}.$$
We want this to be smaller than $\sqrt{M}$ so
\begin{eqnarray*}
\sqrt{M}&\geq& 2zN\sqrt{\frac{M/N(1-M/N)}{k}}\\
M&\geq& 4z^2N^2\frac{M/N(1-M/N)}{k}\\
k&\geq&4z^2N^2\frac{M/N(1-M/N)}{M}\\
k&\geq&4z^2N(1-M/N)
\end{eqnarray*}
so $k=\Omega(N)$ \cite[Exercise 6.13]{nielsen2010quantum}.

It turns out that this is the best bound, in the sense that any classical counting algorithm with a probability at least
3/4 for estimating $M$ correctly to within an accuracy $c\sqrt{M}$ for some constant $c$
must make $\Omega(N)$ oracle calls  \cite[Exercise 6.14]{nielsen2010quantum}, \cite[Table 2.5]{mosca1999quantum}.
So quantum computing gives us a quadratic speedup.

For QWMC, consider the following classical algorithm: take $k$ assignment samples by sampling each bit according
to its normalized weight. For each assignment sample, query the oracle obtaining value $X_i$ with $i=1,\ldots,k$ and estimate the WMC as for the unweighted case: $S=\frac{N}{k}\sum_{i=1}^kX_i$
Variable $Sk/N$ is again binomially distributed with $k$ the number of trials and probability of success
$\widehat{WMC}(\phi,w)$.
In fact, the probability $P(X_i=1)$ is given by
$P(X_i)=\sum_{x}P(X_i,x)=\sum_{x}P(X_i|x)P(x)$
where $P(X_i|x)$ is 1 if $x$ is a model of $\phi$ and 0 otherwise. So
\begin{eqnarray*}
P(X_i)&=&\sum_{x:\phi(x)=1}P(x)=\\
&&\sum_{b_{n}\ldots b_1:\phi(b_{n}\ldots b_1)=1}P(b_{n}\ldots b_1)=\\
&&\sum_{b_{n}\ldots b_1:\phi(b_{n}\ldots b_1)=1}P(b_{n})\ldots P(b_1)=\\
&&\sum_{b_{n}\ldots b_1:\phi(b_{n}\ldots b_1)=1}\prod_{i=1}^{n}\hat w'_{i}=\\
&&\sum_{b_{n}\ldots b_1:\phi(b_{n}\ldots b_1)=1}\prod_{i=1}^{n}\frac{w(b_i)}{V_i}=\\
&&\frac{WMC(\phi,w)}{\prod_{i=1}^{n}V_i}=\widehat{WMC}(\phi,w)
\end{eqnarray*}
This means that we can repeat the reasoning performed with counting: the size of the interval
where the true value of 
$\widehat{WMC}(\phi,w)$ lies is
$$2z\sqrt{\frac{S/N(1-S/N)}{k}}$$
Let us replace $S/N$ by its true value $\widehat{WMC}(\phi,w)$ obtaining
$$2z\sqrt{\frac{\widehat{WMC}(\phi,w)(1-\widehat{WMC}(\phi,w))}{k}}$$
Suppose we want the error below $ 2^{-\lceil\frac{ n}{2}\rceil}$ so
$$2^{-\lceil\frac{ n}{2}\rceil}\geq 2z\sqrt{\frac{\widehat{WMC}(\phi,w)(1-\widehat{WMC}(\phi,w))}{k}} $$

Squaring both members we get (if $n$ is even, if it is odd the result is similar)
$$2^ {-n}\geq 4z^2{\frac{\widehat{WMC}(\phi,w)(1-\widehat{WMC}(\phi,w))}{k}}  $$
and
$$k\geq 4z^22^n\widehat{WMC}(\phi,w)(1-\widehat{WMC}(\phi,w))$$
We want the bound to work for all valules of $\widehat{WMC}(\phi,w)$ and $\widehat{WMC}(\phi,w)(1-\widehat{WMC}(\phi,w)\leq 1/4$ so we must have
$$k \geq z^22^n$$
Therefore $k=\Omega(N)$. This is also the best bound for a classical algorithm, as otherwise we could solve model counting with a better bound than $\Omega(N)$ by setting all weights to 0.5,
So we can conclude that, in the black
box model of computation, estimating the WMC with a probability at least 3/4 and a maximum error for $\widehat{WMC}(\phi,w)$ of $ 2^{-\lceil\frac{ n}{2}\rceil}$ requires  $\Omega(N)$ calls to the oracle for a a classical algorithm.
Therefore QWMC offers a quadratic speedup over classical computation in the black box model.

\section{Conclusion}

We have proposed an algorithm for performing quantum weighted model counting. The algorithm minimally modifies the quantum 
counting algorithm  by just changing the preparation of the state of the second register.
In turn QWMC uses also quantum search, phase estimation and Fourier transform.

Using the black box model of computation, QWMC makes $\Theta(\sqrt{N})$ oracle calls to return
a result whose errors is bounded by $ 2^{-\frac{ n+1}{2}}$ with probability 11/12.
By contrast, the best classical algorithm requires $\Theta(N)$ calls to the oracle.
Thus QWMC offers a quadratic speedup that may be useful in model with high treewidth, where
classical probabilistic inference algorithms have a complexity that is exponential in the treewidth.

\bibliographystyle{aaai}
\bibliography{bibliography/journals_short,bibliography/booktitles_long,bibliography/series_springer,bibliography/series_long,bibliography/publishers_long,bibliography/bibl}

\begin{thebibliography}{}

\bibitem[\protect\citeauthoryear{Aharonov}{1999}]{aharonov1999quantum}
Aharonov, D.
\newblock 1999.
\newblock Quantum computation.
\newblock In {\em Annual Reviews of Computational Physics VI}. World
  Scientific.
\newblock  259--346.

\bibitem[\protect\citeauthoryear{Boyer \bgroup et al\mbox.\egroup
  }{1998}]{boyer1998tight}
Boyer, M.; Brassard, G.; H{\o}yer, P.; and Tapp, A.
\newblock 1998.
\newblock Tight bounds on quantum searching.
\newblock {\em Fortschritte der Physik: Progress of Physics} 46(4-5):493--505.

\bibitem[\protect\citeauthoryear{Brassard, H{\o}yer, and
  Tapp}{1998}]{DBLP:conf/icalp/BrassardHT98}
Brassard, G.; H{\o}yer, P.; and Tapp, A.
\newblock 1998.
\newblock Quantum counting.
\newblock In Larsen, K.~G.; Skyum, S.; and Winskel, G., eds., {\em Automata,
  Languages and Programming, 25th International Colloquium, ICALP'98, Aalborg,
  Denmark, July 13-17, 1998, Proceedings}, volume 1443 of {\em Lecture Notes in
  Computer Science},  820--831.
\newblock Springer.

\bibitem[\protect\citeauthoryear{Chavira and
  Darwiche}{2008}]{DBLP:journals/ai/ChaviraD08}
Chavira, M., and Darwiche, A.
\newblock 2008.
\newblock On probabilistic inference by weighted model counting.
\newblock {\em Artif. Intell.} 172(6-7):772--799.

\bibitem[\protect\citeauthoryear{Cleve \bgroup et al\mbox.\egroup
  }{1998}]{cleve1998quantum}
Cleve, R.; Ekert, A.; Macchiavello, C.; and Mosca, M.
\newblock 1998.
\newblock Quantum algorithms revisited.
\newblock {\em Proceedings of the Royal Society of London. Series A:
  Mathematical, Physical and Engineering Sciences} 454(1969):339--354.

\bibitem[\protect\citeauthoryear{Coppersmith}{2002}]{coppersmith2002approximate}
Coppersmith, D.
\newblock 2002.
\newblock An approximate fourier transform useful in quantum factoring.
\newblock {\em arXiv preprint quant-ph/0201067}.

\bibitem[\protect\citeauthoryear{Darwiche}{2001}]{darwiche2001recursive}
Darwiche, A.
\newblock 2001.
\newblock Recursive conditioning.
\newblock {\em Artif. Intell.} 126(1-2):5--41.

\bibitem[\protect\citeauthoryear{Dechter}{1999}]{dechter1999bucket}
Dechter, R.
\newblock 1999.
\newblock Bucket elimination: A unifying framework for reasoning.
\newblock {\em Artif. Intell.} 113(1-2):41--85.

\bibitem[\protect\citeauthoryear{Gomes, Sabharwal, and
  Selman}{2009}]{DBLP:series/faia/GomesSS09}
Gomes, C.~P.; Sabharwal, A.; and Selman, B.
\newblock 2009.
\newblock Model counting.
\newblock In Biere, A.; Heule, M.; van Maaren, H.; and Walsh, T., eds., {\em
  Handbook of Satisfiability}, volume 185. {IOS} Press.
\newblock  633--654.

\bibitem[\protect\citeauthoryear{Griffiths and
  Niu}{1996}]{griffiths1996semiclassical}
Griffiths, R.~B., and Niu, C.-S.
\newblock 1996.
\newblock Semiclassical fourier transform for quantum computation.
\newblock {\em Physical Review Letters} 76(17):3228.

\bibitem[\protect\citeauthoryear{Grover}{1996a}]{Grover:1996:FQM:237814.237866}
Grover, L.~K.
\newblock 1996a.
\newblock A fast quantum mechanical algorithm for database search.
\newblock In {\em Proceedings of the Twenty-eighth Annual ACM Symposium on
  Theory of Computing}, STOC '96,  212--219.
\newblock New York, NY, USA: ACM Press.

\bibitem[\protect\citeauthoryear{Grover}{1996b}]{grover1996fast}
Grover, L.~K.
\newblock 1996b.
\newblock A fast quantum mechanical algorithm for database search.
\newblock {\em arXiv preprint quant-ph/9605043}.

\bibitem[\protect\citeauthoryear{Grover}{1997}]{grover1997quantum}
Grover, L.~K.
\newblock 1997.
\newblock Quantum mechanics helps in searching for a needle in a haystack.
\newblock {\em Physical review letters} 79(2):325.

\bibitem[\protect\citeauthoryear{Lauritzen and
  Spiegelhalter}{1988}]{lauritzen1988local}
Lauritzen, S.~L., and Spiegelhalter, D.~J.
\newblock 1988.
\newblock Local computations with probabilities on graphical structures and
  their application to expert systems.
\newblock {\em Journal of the Royal Statistical Society: Series B
  (Methodological)} 50(2):157--194.

\bibitem[\protect\citeauthoryear{Mosca}{1999}]{mosca1999quantum}
Mosca, M.
\newblock 1999.
\newblock {\em Quantum computer algorithms}.
\newblock Ph.D. Dissertation, University of Oxford. 1999.

\bibitem[\protect\citeauthoryear{Nielsen and Chuang}{2010}]{nielsen2010quantum}
Nielsen, M., and Chuang, I.
\newblock 2010.
\newblock {\em Quantum Computation and Quantum Information: 10th Anniversary
  Edition}.
\newblock Cambridge University Press.

\bibitem[\protect\citeauthoryear{Pearl}{1988}]{pearl88}
Pearl, J.
\newblock 1988.
\newblock {\em Probabilistic Reasoning in Intelligent Systems: Networks of
  Plausible Inference}.
\newblock Morgan Kaufmann.

\bibitem[\protect\citeauthoryear{Sang, Beame, and
  Kautz}{2005}]{DBLP:conf/aaai/SangBK05}
Sang, T.; Beame, P.; and Kautz, H.~A.
\newblock 2005.
\newblock Performing bayesian inference by weighted model counting.
\newblock In {\em 20th National Conference on Artificial Intelligence},
  475--482.
\newblock Palo Alto, California USA: AAAI Press.

\bibitem[\protect\citeauthoryear{{Shor}}{1994}]{shor}
{Shor}, P.~W.
\newblock 1994.
\newblock Algorithms for quantum computation: discrete logarithms and
  factoring.
\newblock In {\em Proceedings 35th Annual Symposium on Foundations of Computer
  Science},  124--134.
\newblock IEEE Press.

\bibitem[\protect\citeauthoryear{Zhang and
  Poole}{1996}]{DBLP:journals/jair/ZhangP96}
Zhang, N.~L., and Poole, D.~L.
\newblock 1996.
\newblock Exploiting causal independence in {Bayesian} network inference.
\newblock {\em J. Artif. Intell. Res.} 5:301--328.

\end{thebibliography}
\end{document}